\documentstyle[12pt,epsf]{article}
\newlength{\pubnumber} 

\catcode`\@=11
\@addtoreset{equation}{section}

\def\section{\@startsection{section}{1}{\z@}{3.5ex plus 1ex minus .2ex}
 {2.3ex plus .2ex}{\large\bf}}
\def\subsection{\@startsection{subsection}{2}{\z@}{2.3ex plus .2ex}
 {2.3ex plus .2ex}{\bf}}

\def\mbf{\mathbf}

\def\LRS{LRS  }

\def\MP{M_{Pl}}
\def\MS{M_{\rm string}}
\def\GeV{\,{\rm GeV}}
\def\TeV{\,{\rm TeV}}

\begin{document}

\begin{titlepage}
\samepage{
\setcounter{page}{1}
\rightline{ACT--06/01, BU-HEPP--01/02}
\rightline{CTP--TAMU--17/01, OUTP-01--27P}

\rightline{\tt hep-ph/0106060}
\rightline{June 2001}
\vfill
\begin{center}
  {\Large \bf Flat Directions \\in\\ Left--Right Symmetric
 String Derived Models\\}
\vfill
\vfill {\large
Gerald B. Cleaver$^{1,2,3}$\footnote{gcleaver@rainbow.physics.tamu.edu},
David J. Clements$^{4}$\footnote{david.clements@new.oxford.ac.uk}
	and 
Alon E. Faraggi$^{4}$\footnote{faraggi@thphys.ox.ac.uk}}\\
\vspace{.12in}
{\it $^{1}$ Center for Theoretical Physics,
	    Texas A\&M University, College Station, TX 77843\\}
\vspace{.05in}
{\it $^{2}$ Astro Particle Physics Group,
            Houston Advanced Research Center (HARC),\\
            The Mitchell Campus,
            Woodlands, TX 77381\\}
\vspace{.05in}
{\it $^{3}$ Department of Physics,
            Baylor University,
            Waco, TX 76798-7316\\}
\vspace{.05in}
{\it $^{4}$ Theoretical Physics Department, University of Oxford,
            Oxford, OX1 3NP, UK\\}
\vspace{.025in}
\end{center}
\vfill
\begin{abstract}
The only string models known to reproduce the Minimal Supersymmetric 
Standard Model in the low energy effective field theory are those 
constructed in the free fermionic formulation.  
We demonstrate the existence of quasi--realistic free fermionic
heterotic--string models in which supersymmetric singlet flat directions
do not exist. This raises the possibility that supersymmetry is broken
perturbatively in such models by the one--loop Fayet--Iliopoulos term.
We show, however, that supersymmetric flat directions that utilize 
VEVs of some non--Abelian fields in the massless string spectrum
do exist in the model. We argue
that hidden sector
condensates lift the flat directions and break supersymmetry hierarchically.
\end{abstract}
\smallskip}
\end{titlepage}

\setcounter{footnote}{0}

\def\beq{\begin{equation}}
\def\eeq{\end{equation}}
\def\beqn{\begin{eqnarray}}
\def\eeqn{\end{eqnarray}}

\def\no{\noindent }
\def\nolabel{\nonumber }
\def\ie{{\it i.e.}}
\def\eg{{\it e.g.}}
\def\half{{\textstyle{1\over 2}}}
\def\third{{\textstyle {1\over3}}}
\def\quarter{{\textstyle {1\over4}}}
\def\sixth{{\textstyle {1\over6}}}
\def\tenth{{\textstyle {1\over 10}}}
\def\hund{{\textstyle {1\over 100}}}
\def\m{{\tt -}}
\def\p{{\tt +}}

\def\Tr{{\rm Tr}\, }
\def\tr{{\rm tr}\, }

\def\slash#1{#1\hskip-6pt/\hskip6pt}
\def\slk{\slash{k}}
\def\GeV{\,{\rm GeV}}
\def\TeV{\,{\rm TeV}}
\def\y{\,{\rm y}}
\def\SM{Standard--Model }
\def\SUSY{supersymmetry }
\def\SSSM{supersymmetric standard model}
\def\vev#1{\left\langle #1\right\rangle}
\def\l{\langle}
\def\r{\rangle}
\def\o#1{\frac{1}{#1}}

\def\Htw{{\tilde H}}
\def\chibar{{\bar{\chi}}}
\def\qbar{{\bar{q}}}
\def\ibar{{\bar{\imath}}}
\def\jbar{{\bar{\jmath}}}
\def\Hbar{{\bar{H}}}
\def\Qbar{{\bar{Q}}}
\def\abar{{\bar{a}}}
\def\alphabar{{\bar{\alpha}}}
\def\betabar{{\bar{\beta}}}
\def\tautwo{{ \tau_2 }}
\def\thetatwo{{ \vartheta_2 }}
\def\thetathree{{ \vartheta_3 }}
\def\thetafour{{ \vartheta_4 }}
\def\ttwo{{\vartheta_2}}
\def\tthree{{\vartheta_3}}
\def\tfour{{\vartheta_4}}
\def\ti{{\vartheta_i}}
\def\tj{{\vartheta_j}}
\def\tk{{\vartheta_k}}
\def\calF{{\cal F}}
\def\smallmatrix#1#2#3#4{{ {{#1}~{#2}\choose{#3}~{#4}} }}
\def\ab{{\alpha\beta}}
\def\Minv{{ (M^{-1}_\ab)_{ij} }}
\def\bone{{\bf 1}}
\def\bo{{\bf 1}}
\def\ii{{(i)}}
\def\V{{\bf V}}
\def\N{{\bf N}}

\def\bfb{{\bf b}}
\def\bfS{{\bf S}}
\def\bfX{{\bf X}}
\def\bfI{{\bf I}}
\def\ma{{\mathbf a}}
\def\mb{{\mathbf b}}
\def\mS{{\mathbf S}}
\def\mX{{\mathbf X}}
\def\mI{{\mathbf I}}
\def\malpha{{\mathbf \alpha}}
\def\mbeta{{\mathbf \beta}}
\def\mgamma{{\mathbf \gamma}}
\def\mzeta{{\mathbf \zeta}}
\def\mxi{{\mathbf \xi}}

\def\t#1#2{{ \Theta\left\lbrack \matrix{ {#1}\cr {#2}\cr }\right\rbrack }}
\def\C#1#2{{ C\left\lbrack \matrix{ {#1}\cr {#2}\cr }\right\rbrack }}
\def\tp#1#2{{ \Theta'\left\lbrack \matrix{ {#1}\cr {#2}\cr }\right\rbrack }}
\def\tpp#1#2{{ \Theta''\left\lbrack \matrix{ {#1}\cr {#2}\cr }\right\rbrack }}
\def\l{\langle}
\def\r{\rangle}

\def\x#1{\phi_{#1}}
\def\bx#1{{\bar{\phi}}_{#1}}

\def\cl#1{{\cal L}_{#1}} 
\def\bcl#1{\bar{\cal L}_{#1}} 

\def\bt{{\bar 3}}
\def\h#1{h_{#1}}
\def\Q#1{Q_{#1}}
\def\L#1{L_{#1}}

\def\N#1{N_{#1}}
\def\bN#1{{\bar{N}}_{#1}}

\def\S#1{S_{#1}}
\def\Ss#1#2{S_{#1}^{#2}}
\def\bS#1{{\bar S}_{#1}}
\def\Sb#1{{\bar S}_{#1}}
\def\bSs#1#2{{\bar{S}}_{#1}^{#2}}

\def\D#1{D_{#1}}
\def\Ds#1#2{D_{#1}^{#2}}
\def\bD#1{{\bar{D}}_{#1}}
\def\bDs#1#2{{\bar{D}}_{#1}^{#2}}

\def\p#1{\phi_{#1}}
\def\bp#1{{\bar{\phi}}_{#1}}

\def\P#1{\Phi_{#1}}
\def\bP#1{{\bar{\Phi}}_{#1}}
\def\X#1{\Phi_{#1}}
\def\bX#1{{\bar{\Phi}}_{#1}}
\def\Ps#1#2{\Phi_{#1}^{#2}}
\def\bPs#1#2{{\bar{\Phi}}_{#1}^{#2}}
\def\ps#1#2{\phi_{#1}^{#2}}
\def\bps#1#2{{\bar{\phi}}_{#1}^{#2}}
\def\php{\phantom{+}}

\def\H#1{H_{#1}}
\def\bH#1{{\bar{H}}_{#1}}
\def\Hb#1{{\bar{H}}_{#1}}

\def\xH#1#2{H^{#1}_{#2}}
\def\bxH#1#2{{\bar{H}}^{#1}_{#2}}

\def\UA{U(1)_{\rm A}}
\def\QA{Q^{(\rm A)}}
\def\mssm{SU(3)_C\times SU(2)_L\times U(1)_Y} 


\def\inbar{\,\vrule height1.5ex width.4pt depth0pt}

\def\IC{\relax\hbox{$\inbar\kern-.3em{\rm C}$}}
\def\IQ{\relax\hbox{$\inbar\kern-.3em{\rm Q}$}}
\def\IR{\relax{\rm I\kern-.18em R}}
 \font\cmss=cmss10 \font\cmsss=cmss10 at 7pt
\def\IZ{\relax\ifmmode\mathchoice
 {\hbox{\cmss Z\kern-.4em Z}}{\hbox{\cmss Z\kern-.4em Z}}
 {\lower.9pt\hbox{\cmsss Z\kern-.4em Z}}
 {\lower1.2pt\hbox{\cmsss Z\kern-.4em Z}}\else{\cmss Z\kern-.4em Z}\fi}

\def\AEF{A.E. Faraggi}
\def\NPB#1#2#3{{\it Nucl.\ Phys.}\/ {\bf B#1} (#2) #3}
\def\PLB#1#2#3{{\it Phys.\ Lett.}\/ {\bf B#1} (#2) #3}
\def\PRD#1#2#3{{\it Phys.\ Rev.}\/ {\bf D#1} (#2) #3}
\def\PRL#1#2#3{{\it Phys.\ Rev.\ Lett.}\/ {\bf #1} (#2) #3}
\def\PRT#1#2#3{{\it Phys.\ Rep.}\/ {\bf#1} (#2) #3}
\def\MODA#1#2#3{{\it Mod.\ Phys.\ Lett.}\/ {\bf A#1} (#2) #3}
\def\IJMP#1#2#3{{\it Int.\ J.\ Mod.\ Phys.}\/ {\bf A#1} (#2) #3}
\def\nuvc#1#2#3{{\it Nuovo Cimento}\/ {\bf #1A} (#2) #3}
\def\RPP#1#2#3{{\it Rept.\ Prog.\ Phys.}\/ {\bf #1} (#2) #3}
\def\etal{{\it et al\/}}

\hyphenation{su-per-sym-met-ric non-su-per-sym-met-ric}
\hyphenation{space-time-super-sym-met-ric}
\hyphenation{mod-u-lar mod-u-lar--in-var-i-ant}


\setcounter{footnote}{0}
\section{Introduction}
The only string models known to produce the minimal supersymmetric
standard model in the low energy effective field theory are those
constructed in the free fermionic formulation \cite{mshsm}. The
first free fermionic string models that were constructed included
the flipped $SU(5)$ string models \cite{fsu5} (FSU5), the standard--like
string models \cite{fny,slm} (SLM) and the Pati--Salam string models
\cite{alr} (PS). Recently, we constructed such three generation free
fermionic models with the Left--Right Symmetric (LRS) gauge group,
$SU(3)\times SU(2)_L\times SU(2)_R\times U(1)_{B-L}$ \cite{lrsmodels}.
The massless spectrum of the FSU5, SLM and PS free fermionic string
models that have been constructed always contained an ``anomalous''
$U(1)$ symmetry. The anomalous $U(1)_A$ is broken by the
Green--Schwarz--Dine--Seiberg--Witten mechanism \cite{dsw}
in which a potentially large Fayet--Iliopoulos $D$--term
$\xi$ is generated by the VEV of the dilaton field.
Such a $D$--term would, in general, break supersymmetry, unless
there is a direction $\hat\phi=\sum\alpha_i\phi_i$ in the scalar
potential for which $\sum Q_A^i\vert\alpha_i\vert^2<0$ and that
is $D$--flat with respect to all the non--anomalous gauge symmetries 
along with $F$--flat. 
If such a direction
exists, it will acquire a VEV, canceling the Fayet--Iliopoulos
$\xi$--term, restoring supersymmetry and stabilizing the vacuum.
The set of $D$ and $F$ flat constraints is given by,
\beqn
&& \langle D_A\rangle=\langle D_\alpha\rangle= 0;\quad
\langle F_i\equiv
{{\partial W}\over{\partial\eta_i}}\rangle=0\label{dterms}\\
\nonumber\\
&& D_A=\left[K_A+
\sum Q_A^k\vert\chi_k\vert^2+\xi\right]\label{da}\\
&& D_\alpha=\left[K_\alpha+
\sum Q_\alpha^k\vert\chi_k\vert^2\right]~,~\alpha\ne A\label{dalpha}\\
&& \xi={{g^2({\rm Tr} Q_A)}\over{192\pi^2}}M_{\rm Pl}^2
\label{dxi}
\eeqn
where $\chi_k$ are the fields which acquire VEVs of order
$\sqrt\xi$, while the $K$--terms contain fields $\eta_i$
like squarks, sleptons and Higgs bosons whose
VEVs vanish at this scale. $Q_A^k$ and $Q_\alpha^k$ denote the anomalous
and non--anomalous charges,
and $M_{\rm Pl}\approx2\times 10^{18}$ GeV denotes the
reduced Planck mass. The solution ({\it i.e.}\  the choice of fields
with non--vanishing VEVs) to the set of
equations (\ref{dterms})--(\ref{dalpha}),
though nontrivial, is not unique. Therefore in a typical model there exist
a moduli space of solutions to the $F$ and $D$ flatness constraints,
which are supersymmetric and degenerate in energy \cite{moduli}. Much of
the study of the superstring models phenomenology 
(as well as non--string supersymmetric models \cite{savoy})
involves the analysis and classification of these flat directions.
The methods for this analysis in string models have been systematized
in the past few years \cite{systematic,mshsm}. 

It has in general been assumed in the past that in a given string model
there should exist a supersymmetric solution to the $F$ and $D$
flatness constraints. The simpler type of solutions utilize only
fields that are singlets of all the non--Abelian groups in a given
model (type I solutions). More involved solutions (type II solutions),
that utilize also non--Abelian fields, have also been considered
\cite{mshsm}, as well as recent inclusion of non--Abelian fields
in systematic methods of analysis \cite{mshsm}.

In contrast to the case of the FSU5, SLM and PS string models, the
LRS string models \cite{lrsmodels} gave rise to models in which
all the Abelian $U(1)$ symmetries are anomaly free. These models,
therefore, are at a stable point in the moduli space, and the
vacuum remains unshifted. In ref.\  \cite{lrsmodels} we discussed 
the characteristic features of the LRS string models that resulted
in models completely free of Abelian anomalies. On the other hand,
some of the LRS string models that were constructed in ref.
\cite{lrsmodels} did contain an anomalous $U(1)$.
In this paper we examine the supersymmetric flat directions in 
the LRS models that do contain an anomalous $U(1)$ symmetry, 
and find some surprising results. The immediate observation
is that in the LRS string model that we study, there in fact
does not exist a type I solution. Namely, there is no
supersymmetric vacuum that is obtained solely by utilizing
singlet VEVs! Specifically, there is no solution to the 
$D$--term constraints that utilizes solely singlet VEVs!
If a $D$--flat vacuum exists it necessitates the induction of 
a non--vanishing VEV for some non--Abelian fields in the spectrum.
We then show that in fact there exists a $D$--flat solution that
utilizes non--Abelian VEVs.  
While in the past non--Abelian VEVs have been advocated for 
phenomenological considerations \cite{navevs}, this is the
first instance were non--Abelian VEVs are necessitated by
requiring the  existence of a supersymmetric vacuum at the
Planck scale. This is an interesting outcome for the following
reason. It has been argued in the past that NA VEVs necessarily 
induce a mass term in the superpotential that results 
in hierarchical supersymmetry breaking \cite{fh1}. While in the past 
the motivation for the non--Abelian VEVs was purely phenomenological 
\cite{navevs}, here we have an example were the non--Abelian VEVs
are enforced if we require that there is no supersymmetry breaking
VEVs of the Planck scale order. The new features of the LRS
string model may therefore have important bearing on the
issue of supersymmetry breaking. 

Several further comments are in order. First we comment that
the LRS symmetric model under study cannot give rise to a realistic
model for the following reasons. First, all the Higgs doublets
{}from the Neveu--Schwarz sector are projected out by the GSO
projections, which renders the prospect of generating realistic
fermion mass spectrum rather problematic. Second, and more importantly,
the non--Abelian fields $\cl{}$ that are used to cancel the anomalous 
$U(1)$ $D$--term carry fractional electric charge and 
consequently the supersymmetric vacuum state cannot be realistic. 

The new features of the LRS string models have interesting
potential implications from additional perspectives. 
First, as discussed above, the model does not admit 
singlet flat directions. The question then is whether
a given string model always admits a supersymmetric vacuum. 
This has always been assumed in the past, but there is no 
theorem to this effect. In the present model we do show that 
there exists a supersymmetric vacuum at the Planck scale,
but the new observations prompt us to examine the issue
in closely related models which will be reported
in a future publication. However, lets assume that 
a supersymmetric flat direction does not exist. The
implication would be that the vacuum is necessarily 
non--supersymmetric and the supersymmetry breaking 
is of the order of the string scale. However, this breaking
is unlike the breaking that can be induced by projecting
the remaining gravitino with a GSO projection. While in the 
later the supersymmetry breaking is at the level of the spectrum,
in the former the tree level spectrum remains supersymmetric. 
In fact, as a result, also the one--loop partition function,
and hence the one--loop cosmological constant, is vanishing.
One still expects, however, a two--loop contribution to the vacuum 
energy due to the possible non--vanishing $D$--term \cite{sen}. These
new features of the LRS string models may therefore have 
important bearing on the issue of supersymmetry breaking
as well as on that of the vacuum energy. Although we find that
supersymmetric flat directions do exist in the model, we in general
expect that hidden sector condensates break supersymmetry and lift
the flat directions \cite{fh1,mshsm}.
This is unlike the case in type I solutions in which one often
finds solutions that are not lifted by hidden sector condensates.
Thus, the LRS models may reveal a situation in which NA VEVs are
enforced by requiring that the vacuum is supersymmetric at the Planck
scale. Which, in turn, induces hierarchical SUSY breaking
by the hidden sector condensates. In such a situation, the
hierarchical breaking of supersymmetry is no longer a choice,
but is enforced in the vacuum. 

\section{The string model}

The realistic free fermionic string models are constructed
by specifying a set of boundary condition basis vectors
and one--loop GSO projection coefficients \cite{fff}.
The rules for extracting the superpotential terms were
derived in ref. \cite{kln}
The general structure of the models have been discussed in
detail in the past.
The basis vectors which generate the models are divided
into two groups. The first five consist of the NAHE set \cite{nahe}
and are common to all the semi--realistic free fermionic models.
The second consists of three additional boundary condition basis
vectors. Specific details on the construction of the LRS free fermionic
string models are given in ref.\  \cite{lrsmodels}.
  
The left--right symmetric
free fermionic heterotic string model
that we consider is specified in the tables below.
The boundary conditions of the three basis vectors
which extend the NAHE set are shown in Table (\ref{model1}).
Also given in Table (\ref{model1}) are the pairings of 
left-- and right--moving real fermions from the set
$\{y,\omega|{\bar y},{\bar\omega}\}$. These fermions are
paired to form either complex, left-- or right--moving fermions, 
or Ising model operators, which combine a real left--moving fermion with
a real right--moving fermion. The generalized GSO coefficients
determining the physical massless states of Model 1 appear in 
matrix (\ref{phasesmodel1}).
\vskip 0.4truecm

LRS Model 1 Boundary Conditions:
\beqn
 &\begin{tabular}{c|c|ccc|c|ccc|c}
 ~ & $\psi^\mu$ & $\chi^{12}$ & $\chi^{34}$ & $\chi^{56}$ &
        $\bar{\psi}^{1,...,5} $ &
        $\bar{\eta}^1 $&
        $\bar{\eta}^2 $&
        $\bar{\eta}^3 $&
        $\bar{\phi}^{1,...,8} $ \\
\hline
\hline
 ${\malpha}$  &  0 & 0&0&0 & 1~1~1~0~0 & 0 & 0 & 0 &1~1~1~1~0~0~0~0 \\
 ${\mbeta}$   &  0 & 0&0&0 & 1~1~1~0~0 & 0 & 0 & 0 &1~1~1~1~0~0~0~0 \\
 ${\mgamma}$  &  0 & 0&0&0 &
${1\over2}$~${1\over2}$~${1\over2}$~0~0&${1\over2}$&${1\over2}$&${1\over2}$ &
0~1~1~1~1~1~1~0
\end{tabular}
   \nonumber\\
   ~  &  ~ \nonumber\\
   ~  &  ~ \nonumber\\
     &\begin{tabular}{c|c|c|c}
 ~&   $y^3{y}^6$
      $y^4{\bar y}^4$
      $y^5{\bar y}^5$
      ${\bar y}^3{\bar y}^6$
  &   $y^1{\omega}^5$
      $y^2{\bar y}^2$
      $\omega^6{\bar\omega}^6$
      ${\bar y}^1{\bar\omega}^5$
  &   $\omega^2{\omega}^4$
      $\omega^1{\bar\omega}^1$
      $\omega^3{\bar\omega}^3$
      ${\bar\omega}^2{\bar\omega}^4$ \\
\hline
\hline
$\malpha$& 1 ~~~ 1 ~~~ 1 ~~~ 0  & 1 ~~~ 1 ~~~ 1 ~~~ 0  & 1 ~~~ 1 ~~~ 1 ~~~ 0 \\
$\mbeta$ & 0 ~~~ 1 ~~~ 0 ~~~ 1  & 0 ~~~ 1 ~~~ 0 ~~~ 1  & 1 ~~~ 0 ~~~ 0 ~~~ 0 \\
$\mgamma$& 0 ~~~ 0 ~~~ 1 ~~~ 1  & 1 ~~~ 0 ~~~ 0 ~~~ 0  & 0 ~~~ 1 ~~~ 0 ~~~ 1 \\
\end{tabular}
\label{model1}
\eeqn

LRS Model 1 Generalized GSO Coefficients:
\begin{equation}
{\bordermatrix{
          &{\bf 1}&\mS & &{\mb_1}&{\mb_2}&{\mb_3}& &{\malpha}&{\mbeta}&{\mgamma}\cr
       {\bf 1}&~~1&~~1 & & -1   &  -1 & -1  & & ~~1     & ~~1   & ~~i   \cr
           \mS&~~1&~~1 & &~~1   & ~~1 &~~1  & &  -1     &  -1   &  -1   \cr
	      &   &    & &      &     &     & &         &       &       \cr
       {\mb_1}& -1& -1 & & -1   &  -1 & -1  & &  -1     &  -1   & ~~i   \cr
       {\mb_2}& -1& -1 & & -1   &  -1 & -1  & &  -1     &  -1   & ~~i   \cr
       {\mb_3}& -1& -1 & & -1   &  -1 & -1  & &  -1     &  -1   & ~~i   \cr
	      &   &    & &      &     &     & &         &       &       \cr
     {\malpha}&~~1& -1 & &~~1   & ~~1 &~~1  & & ~~1     & ~~1   & ~~1   \cr
      {\mbeta}&~~1& -1 & & -1   &  -1 &~~1  & &  -1     &  -1   &  -1   \cr
     {\mgamma}&~~1& -1 & &~~1   &  -1 &~~1  & &  -1     &  -1   & ~~1   \cr}}
\label{phasesmodel1}
\end{equation}

In matrix (\ref{phasesmodel1}) only the entries above the diagonal are
independent and those below and on the diagonal are fixed by
the modular invariance constraints. 
Blank lines are inserted to emphasize the division of the free
phases between the different sectors of the realistic
free fermionic models. Thus, the first two lines involve
only the GSO phases of $c{{\{{\bf 1},\mS\}}\choose \ma_i}$. The set
$\{{\bf 1},\mS\}$ generates the $N=4$ model with $\mS$ being the
space--time supersymmetry generator and therefore the phases
$c{\mS\choose{\ma_i}}$ are those that control the space--time supersymmetry
in the superstring models. Similarly, in the free fermionic
models, sectors with periodic and anti--periodic boundary conditions,
of the form of $\mb_i$, produce the chiral generations.
The phases $c{\mb_i\choose \mb_j}$ determine the chirality
of the states from these sectors. 

We note that the boundary condition basis vectors that generate the string 
model are those of Model 3 of ref.\  \cite{lrsmodels}. The 
two models differ in the GSO phase $c{{\mb_3}\choose \mbeta}$,
with $c{{\mb_3}\choose \mbeta}=-1$ in the string model above and 
$c{{\mb_3}\choose \mbeta}=+1$ in Model 3 of \cite{lrsmodels}.
As we elaborate below, the consequence of this GSO phase change
is that the gauge symmetry is enhanced, with one of the Abelian generators
being absorbed into the enhanced non--Abelian gauge symmetry.
Consequently, the number of Abelian group factors is reduced,
which simplifies somewhat the analysis of the $D$--flat directions.
However, the results that we discuss here are independent of this
simplification and therefore also hold in Model 3 of ref.\  \cite{lrsmodels}. 

The final gauge group of the string model arises
as follows: In the observable sector the NS boundary conditions 
produce gauge group generators for 
\beq
SU(3)_C\times SU(2)_L\times SU(2)_R\times U(1)_C\times U(1)_{1,2,3}\times
U(1)_{4,5,6}
\eeq
Thus, the $SO(10)$ symmetry is broken to
$SU(3)\times SU(2)_L\times SU(2)_R\times U(1)_C$,
where, 
\begin{equation}
U(1)_C={\rm Tr}\, U(3)_C~\Rightarrow~Q_C=
			 \sum_{i=1}^3Q({\bar\psi}^i).
\label{u1c}
\end{equation}
The flavor $SO(6)^3$ symmetries are broken to $U(1)^{3+n}$ with
$(n=0,\cdots,6)$. The first three, denoted by $U(1)_{j}$ $(j=1,2,3)$, arise 
{}from the world--sheet currents ${\bar\eta}^j{\bar\eta}^{j^*}$.
These three $U(1)$ symmetries are present in all
the three generation free fermionic models which use the NAHE set. 
Additional horizontal $U(1)$ symmetries, denoted by $U(1)_{j}$ 
$(j=4,5,...)$, arise by pairing two real fermions from the sets
$\{{\bar y}^{3,\cdots,6}\}$, 
$\{{\bar y}^{1,2},{\bar\omega}^{5,6}\}$, and
$\{{\bar\omega}^{1,\cdots,4}\}$. 
The final observable gauge group depends on
the number of such pairings. In this model there are the 
pairings, ${\bar y}^3{\bar y}^6$, ${\bar y}^1{\bar\omega}^5$
and ${\bar\omega}^2{\bar\omega}^4$, which generate three additional 
$U(1)$ symmetries, denoted by $U(1)_{4,5,6}$\footnote{It is 
important to note that the existence of these three additional 
$U(1)$ currents is correlated with a superstringy doublet--triplet
splitting mechanism \cite{ps}. Due to these extra $U(1)$ symmetries 
the color triplets from the NS sector are projected out of the spectrum 
by the GSO projections while the electroweak doublets remain in the 
light spectrum.}. 

In the hidden sector, 
which arises from the complex
world--sheet fermions ${\bar\phi}^{1\cdots8}$,
the NS boundary conditions produce the generators of
\beq
SU(3)_{H_1}\times U(1)_{H_1}\times U(1)_{7^\prime}\times
SU(3)_{H_2}\times U(1)_{H_2}\times U(1)_{8^\prime}\, .
\eeq
$U(1)_{H_1}$ and $U(1)_{H_2}$
correspond to the combinations of the world--sheet charges
\begin{eqnarray}
Q_{H_1}&=&Q({\bar\phi^1})-Q({\bar\phi^2})-Q({\bar\phi^3})+Q({\bar\phi^4})-
\sum_{i=5}^7Q({\bar\phi})^i+Q({\bar\phi})^8,\label{qh1}\\
Q_{H_2}&=&\sum_{i=1}^4Q({\bar\phi})^i -Q({\bar\phi^5})+
\sum_{i=6}^8Q({\bar\phi})^i.
\label{qh2}
\end{eqnarray}

The sector $\mzeta\equiv1+\mb_1+\mb_2+\mb_3$ produces the representations 
$(3,1)_{-5,0}\oplus({\bar3},1)_{5,0}$ and 
$(1,3)_{0,-5}\oplus(1,{\bar3})_{0,5}$
of 
$SU(3)_{H_1}\times U(1)_{H_1}$ and $SU(3)_{H_2}\times U(1)_{H_2}$.
Thus, the $E_8$ symmetry
reduces to 
$SU(4)_{H_1}\times SU(4)_{H_2}\times U(1)^2$. 
The additional $U(1)$'s in $SU(4)_{H_{1,2}}$ are given by the
combinations in eqs.~(\ref{qh1}) and (\ref{qh2}), respectively.
The remaining $U(1)$ symmetries in the
hidden sector, $U(1)_{7^\prime}$ and $U(1)_{8^\prime}$,
correspond to the combination of world--sheet charges
\begin{eqnarray}
Q_{7^\prime}&=&Q({\bar\phi^1})-Q({\bar\phi^7}),\label{q7prime}\\
Q_{8^\prime}&=&Q({\bar\phi^1})-\sum_{i=2}^4Q({\bar\phi})^i +
\sum_{i=5}^7Q({\bar\phi})^i+Q({\bar\phi^8}).
\label{q8prime}
\end{eqnarray}

In addition to the NS and $\mzeta$ sector the string model contains
a combination of non--NAHE basis vectors with $\mX_L\cdot \mX_L=0$,
which therefore may give rise to additional space--time vector bosons.
The vector combination is given by $\mX\equiv\mzeta+2\mgamma$, where
$\mzeta\equiv1+\mb_1+\mb_2+\mb_3$. This combination arises only from
the NAHE set basis vectors plus $2\mgamma$, with $\gamma$ inducing the
left--right symmetry breaking pattern $SO(6)\times SO(4)\rightarrow
SU(3)\times U(1)\times SU(2)_L\times SU(2)_R$, and is independent of
the assignment of periodic boundary conditions in the basis vectors
$\malpha$, $\mbeta$ and $\mgamma$. This vector combination is therefor
generic for the pattern of symmetry breaking $SO(10)\rightarrow
SU(3)_C\times U(1)_C\times SU(2)_L\times SU(2)_R$, in NAHE based models. 

The sector $\mX$ 
gives rise to six additional space--time vector bosons 
which are charged with respect to the world--sheet
$U(1)$ currents, and transform as $3\oplus{\bar 3}$
under $SU(3)_C$. 
These additional gauge bosons enhance the $SU(3)_C\times U(1)_{C^\prime}$
symmetry to  $SU(4)_C$, where $U(1)_{C^\prime}$ is given by
the combination of world--sheet charges,
\begin{equation}
Q_{C^\prime}=Q({\bar\psi^1})-Q({\bar\psi^2})-Q({\bar\psi^3})-
\sum_{i=1}^3Q({\bar\eta})^i+Q({\bar\phi}^7)-Q({\bar\phi}^8).
\label{u1su4}
\end{equation}
The remaining orthogonal $U(1)$ combinations are 
\begin{eqnarray}
Q_{1^\prime}&=&~~Q_1-Q_2,\nonumber\\
Q_{2^\prime}&=&~~Q_1+Q_2-2Q_3,\nonumber\\
Q_{3^\prime}&=&3Q_C-(Q_1+Q_2+Q_3),\nonumber\\
Q_{7^{\prime\prime}}&=&Q_C+3(Q_1+Q_2+Q_3)+5Q_{7^\prime}.
\label{u1com}
\end{eqnarray}
and $Q_{4,5,6,8^\prime}$ are unchanged.
Thus, the full massless spectrum transforms under the final gauge group, 
$SU(4)_C\times SU(2)_L\times SU(2)_R\times 
U(1)_{1^\prime,2^\prime,3^\prime}\times U(1)_{4,5,6}\times 
SU(4)_{H_1}\times SU(4)_{H_2}\times U(1)_{7^{\prime\prime},8^{\prime}}$.

In addition to the graviton, 
dilaton, antisymmetric sector and spin--1 gauge bosons, 
the NS sector gives 
three pairs of $SO(10)$ singlets with 
$U(1)_{1,2,3}$ charges; and three singlets of the entire four 
dimensional gauge group.

The states from the sectors $\mb_j\oplus \mb_j+\mX~(j=1,2,3)$
produce the three light generations. The states from these sectors and their
decomposition under the entire gauge group are shown in Table 1 of the 
Appendix. 
The leptons (and quarks) are singlets of the color $SU(4)_{H_1,H_2}$ 
gauge groups and 
the $U(1)_{8^{\prime}}$ symmetry of eq.~(\ref{q8prime}) 
becomes a gauged leptophobic symmetry.
The leptophobic $U(1)$ symmetry arises from a combination of the $U(1)_{B-L}$ 
symmetry with a family universal combination of the flavor and hidden
$U(1)$ symmetries \cite{masip}. The remaining massless states in the model
and their quantum numbers are also given in Table 1.

We next turn to the definition of the weak--hypercharge 
in this \LRS model. Due to the enhanced
symmetry there are several possibilities to define a weak--hypercharge
combination which is still family universal and reproduces 
the correct charge assignment for the Standard Model fermions. 
One option is to define the weak--hypercharge
with the standard $SO(10)$ embedding, as in eq.~(\ref{U1Y}),
\beq
          U(1)_Y~=~{1\over3}\,U(1)_C~+~{1\over2}\,U(1)_L~.
\label{U1Y}
\eeq
This is identical to the weak--hypercharge definition in 
$SU(3)\times SU(2)\times U(1)_Y$  free fermionic models, which
do not have enhanced symmetries, as for example in Model 3 of
ref.\  \cite{lrsmodels}. The weak hypercharge definition of eq. 
(\ref{U1Y}) reproduces the canonical MSSM normalization of the
weak hypercharge, $k_Y=5/3$.
Alternatively, we can define the weak--hypercharge to be the combination
\beq
         U(1)_Y~=~{1\over2}\,U(1)_L~-~ {1\over{10}}(U(1)_{3^\prime}~+~
{1\over3} U_{7^{\prime\prime}})
\label{U1Y2}
\eeq
where $U(1)_{3^\prime}$ and $U(1)_{7^{\prime\prime}}$ 
are given in (\ref{u1com}). This combination still reproduces the
correct charge assignment for the Standard Model states.
The reason being that the states from the sectors $\mb_i$ $i=1,2,3$
which are identified with the Standard Model states, are not charged
with respect to the additional Cartan subgenerators that form the modified
weak hypercharge definition. 
In some models it is found that such alternative definitions
allow all massless exotic states to be integrally charged. The 
price, however, is that the Ka\v{c}--Moody level of the
weak hypercharge current as defined in eq. (\ref{U1Y2})
is no longer the canonical $SO(10)$ normalization and the 
simple unification picture is lost. We conclude that the model
admits a sensible weak--hypercharge definition, which for our
purpose here is sufficient. We stress
that our objective here is not to present the \LRS string model
as a semi--realistic model, but rather to study the new features
pertaining to the existence of a supersymmetric vacuum.

\section{Anomalous $U(1)$}\label{anomalousu1}

The string model contains an anomalous $U(1)$ symmetry. The anomalous
$U(1)$ is a combination of $U(1)_4$, $U(1)_5$ and $U(1)_6$, which are
generated by the world--sheet complex fermions ${\bar y}^3{\bar y}^6$,
${\bar y}^1{\bar\omega}^5$ and ${\bar\omega}^2{\bar\omega}^4$, respectively.
The anomalous $U(1)$ that arises in this model is therefore of a
different origin than the one that typically arises in the 
FSU5, SLM and PS free fermionic string models. The difference 
between the two cases is discussed in detail in ref.\  \cite{lrsmodels}. 
In short, the main distinction is that in the case of the 
FSU5, SLM and PS string models, the $U(1)$ symmetries, $U(1)_1$,
$U(1)_2$ and $U(1)_3$ which are embedded in the observable $E_8$
are necessarily anomalous because of the symmetry breaking pattern
$E_6\rightarrow SO(10)\times U(1)_A$ \cite{cf1}, whereas in the LRS models
they are necessarily anomaly free because the models do not admit
the pattern $E_6\rightarrow SO(10)\times U(1)_A$. Consequently,
in the LRS string models the anomalous $U(1)$ can only arise from 
$U(1)$ currents that arise from the six dimensional internal 
manifold, rather than from the $U(1)$ currents of the observable
$E_8$. This distinction, as we demonstrate further below, is
potentially important because typically the non--Abelian
singlets that arise from the untwisted sector, and are used
to cancel the $U(1)_A$ $D$--term equation, are charged with respect
to $U(1)_{1,2,3}$, but not with respect to $U(1)_{4,5,6}$, which
arise from the internal ``compactified'' degrees of freedom.

The anomalous $U(1)$ generates a Fayet--Iliopoulos $D$--term, which breaks
supersymmetry and destabilizes the vacuum. Stabilization of the vacuum 
implies that the vacuum is shifted by a VEV which cancels the anomalous
$U(1)$ $D$--term and restores supersymmetry. If such a direction
in the scalar potential does not exist, it would imply that supersymmetry
is necessarily broken and a supersymmetric vacuum does not exist. 
Additionally a supersymmetric vacuum also requires that $F$--flatness
is also respected in the vacuum. 
The anomalous $U(1)_A$ combination is given by
\beq
U_A\equiv U_4+U_5+U_6,
\label{anomau1infny}
\eeq
with ${\rm Tr}Q_A=-72$.
The two orthogonal linear combinations,
\beqn
U_{4^\prime} &=& U_4-U_5 \\
U_{5^\prime} &=& U_4+U_5-2 U_6 
\nolabel
\eeqn
are both traceless.

Since ${\rm Tr}Q_A<0$, the sign for the Fayet--Iliopoulos term is negative.
Requiring $D$--flatness then implies that that there must exist a direction
in the scalar potential in which a field (or a combination of fields)
with positive total $U(1)_A$ charge, gets a VEV and cancels the $U(1)_A$
$D$--term. Examining the massless spectrum of the model, given in Table 1,
we immediately note that the model does not contain any non--Abelian
singlet fields with such a charge. Therefore, if a supersymmetric
vacuum exists, some non--Abelian fields must get a VEV. From Table 1
it is seen that the only states that carry positive $U(1)_A$ charge
are the $SU(2)_L$ and $SU(2)_R$ doublets from the three
sectors $\mb_k + \mzeta + 2 \mgamma \equiv \bo+\mb_i+\mb_j+ 2 \mgamma$,  
$(i,j,k=1,2,3)$ with $i$, $j$, $k$ all distinct.
This then implies that $SU(2)_L$ or $SU(2)_R$ 
must be broken in the vacuum. The same result holds also in
Model 3 of ref.\  \cite{lrsmodels}, in which there is no
gauge enhancement from the sector $\mzeta+2\mgamma$.
We note, however, that the doublets from these sectors
carry fractional $\pm1/2$ charge with respect to electric charge as
defined in Eq. (\ref{U1Y}). While there exist
alternative  definitions of the weak--hypercharge that allow these
states to be integrally charged, the primary question of interest here
is whether a supersymmetric vacuum exist at all!
In the model under consideration this is contingent on
finding $D$--flat direction, which are also $F$--flat.
We next turn to examine whether a $D$-flat direction exist in this model. 

\section{$D$--Flat Directions}

\def\P#1{\Phi_{#1}}
\def\Pp#1{\Phi^{'}_{#1}}
\def\Pb#1{{\bar{\Phi}}_{#1}}
\def\bP#1{{\bar{\Phi}}_{#1}}
\def\Pbp#1{{\bar{\Phi}}^{'}_{#1}}
\def\Ppb#1{{\bar{\Phi}}^{'}_{#1}}
\def\Ppx#1{\Phi^{(')}_{#1}}
\def\Pbpx#1{\bar{\Phi}^{(')}_{#1}}

\def\p#1{\phi_{#1}}
\def\pp#1{\phi^{'}_{#1}}
\def\pb#1{{\bar{\phi}}_{#1}}
\def\bp#1{{\bar{\phi}}_{#1}}
\def\pbp#1{{\bar{\phi}}^{'}_{#1}}
\def\ppb#1{{\bar{\phi}}^{'}_{#1}}
\def\ppx#1{\phi^{(')}_{#1}}
\def\pbpx#1{\bar{\phi}^{(')}_{#1}}

\def\pbx#1{{\bar{\phi}_{#1 +}}}
\def\px#1{{     {\phi}_{#1 +}}}
\def\pbn#1{{\bar{\phi}_{#1 -}}}
\def\pn#1{{     {\phi}_{#1 -}}}
\def\Hp{H^{a}_{+}}
\def\Hm{H^{a}_{-}}
\def\Hbp{\bar{H}_{+}}
\def\Hbm{\H^{b}_{-}}
\def\cH#1{{\cal H}_{#1}}
\def\cHb#1{{\bar{\cal H}}_{#1}}
\def\cL#1{{\cal L}_{#1}}

\def\Db#1{{\bar{D}}_{#1}}
\def\Dm#1{{\mathbf D}_{#1}}
\def\Dmv#1{{\mathbf D}^{v}_{#1}}

In Tables 1 and 2 we have listed all of the massless states that appear in
our LRS string model. There are a total of 68 fields, 38 of which may be used to
form 19 sets of vector--like pairs of fields. 
Of these 19 vector--like pairs, 13 pairs are singlets under all non--Abelian 
gauge groups, 
while three pairs are $-\mathbf 4$/$\mathbf 4$ sets
under $SU(4)_C$ and two pairs are $\mathbf 6$/$\mathbf 6$'s sets under
$SU(4)_{H_1}$. The 30 non--vector--like fields are all non--Abelian reps. 
That is, all singlets occur in vector--like pairs.

The anomalous charge trace of $U(1)_A$ is negative for the LRS
Model 1.  Thus, the anomaly can only be cancelled by fields
with positive anomalous charge. In this model  
{\it none} of the NA singlets carry anomalous $\QA$.
The only fields with positive anomalous charge are
three $SU(2)_L$ doublets, $\cL{L1,L2,L3}$ 
and three $SU(2)_R$ doublets, $\cL{R1,R2,R3}$. Hence
any possible flat solutions automatically break the initial
$SU(2)_L \times SU(2)_R$ symmetry to a subgroup. 
Good phenomenology would clearly prefer the subgroup 
$SU(2)_L \times U(1)_R$.

Since we have clearly shown that no non--Abelian singlet $D$--flat directions
are possible for this model, it is possible that no $D$--flat directions
exist. That is, the severity of non--Abelian $D$--flat constraints may allow 
for no solutions. Thus, this model either 
(i) automatically breaks $SU(2)_L \times SU(2)_R$ 
or
(ii) has no flat directions and, therefore, breaks supersymmetry at
the string scale. In either case, this is a very interesting model.

To systematically study $D$--flatness for this model, we first
generate a complete basis of directions $D$--flat for all non--anomalous
Abelian symmetries. We provide this basis in Table 2 of the Appendix.
For a given row in Table 2, the first column entry denotes the name
of the $D$--flat basis direction. The next row specifies the anomalous
charge of the basis direction. The following seven entries specify the
ratios of the norms of the VEVs of the fields common to these directions. 
The first five of these fields have vector--like partners. For these, a
negative norm indicates the vector--partner acquires the VEV, rather than
the field specified at the top of the respective column. The last two of
these seven fields are not vector--like. Thus, the norm must be non--negative
for each of these for a flat direction formed from a linear combination of
basis directions to be physical. The next to last entry specifies the norm
of the VEV of the field unique to a given basis direction, while the
identity of the unique field is given by the last entry.

We have labeled these $D$--flat directions as $\Dm{1}$ through $\Dm{41}$.
The first eight $D$--flat directions ($\Dm{1}$ to $\Dm{8}$) carry a positive
net anomalous charge. The next fourteen ($\Dm{9}$ to $\Dm{22}$) carry a
negative net anomalous charge, while the remaining nineteen ($\Dm{23}$ to
$\Dm{41}$) lack a net anomalous charge. There are two classes of basis
vectors lacking anomalous charge. The first class contains six directions
for which the unique field is non--vector--like. These directions also contain
VEVs for $\cH{1}$ $(\cHb{1})$ and/or $\cH{2}$ $(\cHb{2})$. The second class
contains thirteen basis directions wherein the unique fields are vector--like
and which do not contain $\bH{2'}$ and/or $\bH{4'}$. Thus, these thirteen
directions are themselves vector--like and are denoted as such by a
superscript ``$v$''. For each vector--like basis direction, $\Dmv{}$ there
is a corresponding $-\Dmv{}$ direction, wherein the fields in $\Dmv{}$ 
are replaced by their respective vector--like partners. 

None of the positive $\QA$ directions are good in themselves because 
one or both of $|\vev{\bH{2'}}|^2$ and $|\vev{\bH{4'}}|^2$ are non--zero
and negative while $|\vev{\bH{2'}}|^2$ and $|\vev{\bH{4'}}|^2$ are not
vector--like reps. In particular, the $\QA= 12$ directions have either 
    $|\vev{\bH{2'}}|^2= -2$ and 
    $|\vev{\bH{4'}}|^2=  0$ or 
    $|\vev{\bH{2'}}|^2=  0$ and 
    $|\vev{\bH{4'}}|^2= -2$, 
    while the $\QA= 24$ directions all have 
$|\vev{\bH{2'}}|^2= |\vev{\bH{4'}}|^2= -4$. 
In this basis we also find that the $|\vev{\bH{2'}}|^2$ and $|\vev{\bH{4'}}|^2$
charges of all of the $\QA=0$ directions are zero or negative.
So the $|\vev{\bH{2'}}|^2$ and $|\vev{\bH{4'}}|^2$ negative charges on the
positive $\QA$ directions cannot be made zero or positive by adding $\QA=0$
directions to $\QA >0$ directions. In contrast, all of the $\QA= -12$
directions have either
    $|\vev{\bH{2'}}|^2= 2$ and 
    $|\vev{\bH{4'}}|^2= 0$ or 
    $|\vev{\bH{2'}}|^2= 0$ and 
    $|\vev{\bH{4'}}|^2= 2$;
    the $\QA= 24$ directions all have either
    $|\vev{\bH{2'}}|^2= |\vev{\bH{4'}}|^2= 4$ or 
    $|\vev{\bH{2'}}|^2= |\vev{\bH{4'}}|^2= 2$; while the       
$\QA= -48$ directions all have 
$|\vev{\bH{2'}}|^2= |\vev{\bH{4'}}|^2= 4$ .
Therefore, physical $D$--flat directions must necessarily be formed from
linear combinations of $\QA> 0$ and $\QA<0$ directions
such that the net $\QA$,  $|\vev{\bH{2'}}|^2$, and $|\vev{\bH{4'}}|^2$ 
are all positive. Physical $D$--flat directions may also contain 
$\QA= 0$ components that keep $|\vev{\bH{2'}}|^2$, $|\vev{\bH{4'}}|^2 \ge 0$.

The specific values of $\QA$, $|\vev{\bH{2'}}|^2$, and $|\vev{\bH{4'}}|^2$ 
in the basis directions 
indicate that the roots of all physical flat directions 
must contain either $\Dm{19}$ or $\Dm{20}$ and combinations of
basis vectors $\Dm{1}$, $\Dm{2}$, $\Dm{3}$, and $\Dm{4}$ of the form
\beq
n_1\Dm{1} + n_2\Dm{2} + n_3\Dm{3} + n_4\Dm{4} 
+ n_{19}\Dm{19} + n_{20} \Dm{20},
\label{sol1a}
\eeq
where the non--negative integers  
$n_1$, $n_2,$ $n_3$, $n_4$, $n_{19}$, $n_{20}$ satisfy the constraints
\beqn
 n_1 + n_2 + n_3 + n_4 - 2 n_{19} - 2 n_{20} &  >& 0; \label{sol1b1}\\
-n_1 - n_2             + 2 n_{19} + 2 n_{20} &\ge& 0; \label{sol1b2}\\
-n_3 - n_4             + 2 n_{19} + 2 n_{20} &\ge& 0. \label{sol1b3}
\eeqn
For example, one of the simplest $D$--flat solutions for all Abelian
gauge groups 
is $n_1= n_2=n_3=n_4=2$, $n_{19}= n_{20}= 1$. This direction is simply
$|\vev{\cL{L1}}|^2 = |\vev{\cL{L2}}|^2 = |\vev{\cL{L3}}|^2   
=|\vev{\cL{R1}}|^2 = |\vev{\cL{R2}}|^2 = |\vev{\cL{R3}}|^2 $.
The corresponding fields are three exotic $SU(2)_L$ doublets,
$\cL{L1}$, $\cL{L2}$, and $\cL{L3}$,
and three exotic $SU(2)_R$ doublets,
$\cL{R1}$, $\cL{R2}$, and $\cL{R3}$. 
These six fields are singlets under all other non--Abelian groups.  

For this model any $D$--flat direction must contain
$SU(2)_L$ or $SU(2)_R$ fields.
Thus, let us examine more closely $SU(2)$ $D$--flat 
constraints. The only $SU(2)$ fields in this model are doublet
representations, which we generically denote $L_i$.
Thus, the related three $SU(2)$ $D$--terms, 
\beqn
D_{a=1,2,3}^{SU(2)}&\equiv& \sum_m L_i^{\dagger}
T^{SU(2)}_{a=1,2,3} L_i\,\, , 
\label{dtgen} 
\eeqn
contain matrix generators $T^{SU(2)}_a$ that 
take on the values of the three Pauli matrices,
\beq
\sigma_x =
\left (                    
\begin{array}{cc}
0 & 1 \\
1 & 0 \\
\end{array} \right ), \,\,
\sigma_y =
\left (
\begin{array}{cc}
0 & -i \\
i &  0 \\
\end{array} \right ), \,\,
\sigma_z =
\left (
\begin{array}{cc}
1 & 0 \\
0 & -1 \\
\end{array} \right ),
\label{pauli}
\eeq
respectively.

As discussed in \cite{nap} and \cite{cfnw}, 
each component of the vector $\vec{D}^{SU(2)}$ is the total
``spin expectation value'' in the given direction of the internal space, 
summed over all $SU(2)$ doublet fields of the gauge group.  
Thus, for all of the $\vev{D^{SU(2)}_a}$ to 
vanish, the $SU(2)$ VEVs must be chosen such that the total
$\hat{x}, \hat{y},$ and $\hat{z}$ expectation values are zero.
Abelian $D$--flatness constraints 
{}from any  extra $U(1)$ charges carried by the doublet
generally restrict 
the normalization length, $L_i^{\dagger}L_i$, of a ``spinor'' $L_i$ 
to integer units. 
Thus, since 
each spinor has a length and direction associated with it, 
$D$--flatness requires the sum, placed tip--to--tail, to be zero.
Let us choose for an explicit representation 
of a generic $SU(2)$ doublet $L(\theta, \phi)$
that used in \cite{nap}:  
\beqn
L(\theta, \phi) \equiv
A \left ( 
\begin{array}{c}
\cos{\frac{\theta}{2}} \, e^{-i \frac{\phi}{2}} \\
\sin{\frac{\theta}{2}} \, e^{+i \frac{\phi}{2}} \\
\end{array} \right )\, ,
\label{spinor}
\eeqn
where $A$ is the overall amplitude of the VEV.
The range of physical angles, 
$\theta = 0 \rightarrow \pi$ and $\phi = 0 \rightarrow 2\pi$
provide for the most general possible doublet.
(Note the $\phi$ phase freedom for $\theta = 0,\pi$.) 
Each such $\theta$, $\phi$ combination
carries a one--to--one geometrical correspondence.  

The contribution of  $L(\theta, \phi)$ to each $SU(2)$ $D$--term is,
\beqn
D_{1}^{SU(2)}(L) &\equiv&  L^{\dagger}\,  \left (                    
\begin{array}{cc}
0 & 1 \\
1 & 0 \\
\end{array} \right )\, 
L =  |A|^2 \sin\, \theta \,\, \cos\, \phi \label{dl1}\\ 
D_{2}^{SU(2)}(L) &\equiv&  L^{\dagger}\,  \left (                    
\begin{array}{cc}
0 & - i \\
i &  0 \\
\end{array} \right )\, 
L =  |A|^2 \sin\, \theta \,\, \sin\, \phi \label{dl2}\\
D_{3}^{SU(2)}(L) &\equiv&  L^{\dagger}\,  \left (                    
\begin{array}{cc}
1 &   0 \\
0 &  -1 \\
\end{array} \right )\,  L =  |A|^2 \cos\, \theta\,\, . \label{dl3}
\eeqn
A doublet's $D$-term contribution for any extra $U(1)$
charges carried by it is, 
\beqn
D_{U(1)}(L) &\equiv&  Q^{U(1)} |L|^2 =  Q^{U(1)} |A|^2  \label{dlu}\,\, .
\eeqn
{}From this we can see that the VEVS of three $SU(2)$ doublets $\cL{i=1,2,3}$
with equal norms $|A_1|^2=|A_2|^2=|A_3|^2\equiv |A|^2$  
can, indeed, produce an $SU(2)$ $D$--flat direction with 
the choice of angles (in radians), $\theta_1 = 0$, $\theta_2 =
\theta_3 = 2\pi/3$, $\phi_2 = 0$, $\phi_3 = \pi$.
For these angles, the $SU(2)$ $D$--vectors added tip--to--tail
for the three doublets form an equilateral triangle with starting
and ending points at the origin. In other words, the total $\hat x$,
$\hat y$, and $\hat z$ expectation values are all zero!

This flat direction gives a specific example of what will occur
for every flat direction of this model: non--Abelian VEVs (for at
least $SU(2)_L$ or $SU(2)_R$ doublets) are a necessary effect of
the retention of spacetime supersymmetry. This has profound implications
for this model! First, $SU(2)_L\times SU(2)_R$ gauge symmetry is
automatically broken at the FI scale. Second, non--Abelian condensates 
necessarily form.

Let us consider (ignoring effects of other possible field condensates 
that might develop significantly below the FI--scale) 
the status of $F$--flatness for the particular flat direction 
$|\vev{\cL{L1}}|^2 = |\vev{\cL{L2}}|^2 = |\vev{\cL{L3}}|^2   
=|\vev{\cL{R1}}|^2 = |\vev{\cL{R2}}|^2 = |\vev{\cL{R3}}|^2$.
{}From gauge invariance arguments, we find that $F$--flatness remains
to all finite order. No related dangerous terms appear in the
superpotential. We do however expect dangerous $F$--breaking terms to
appear at finite orders for all but a few of the more complicated 
$D$--flat directions.

For a generic $SU(N_c)$ gauge group containing $N_f$ flavors 
of matter states in vector--like pairings 
$H_i {\bar H}_i$, $i= 1,\, \dots\, N_f$,
the gauge coupling $g_s$, 
though weak at the string scale $\MS$, becomes strong
for $N_f < N_c$ at a condensation scale defined by 
\beqn
\Lambda = \MP {\rm e}^{8 \pi^2/\beta g_s^2}\, ,
\label{consca}
\eeqn
where the $\beta$--function is given by,
\beqn
\beta = - 3 N_c + N_f\, .
\label{befn}
\eeqn
The $N_f$ flavors counted are only those that ultimately receive 
masses $m\ll \Lambda$.

Our model contains 
three vector--like ${\mbf 4}-{\bar {\mbf 4}}$ pairs
for each hidden sector $SU(4)_{H_{1,2}}$ gauge group, along with 
an additional ${\mbf 6}-{\mbf 6}$ pair for $SU(4)_{H_2}$.  
We have computed all possible mass terms for the non--Abelian fields
resulting from our simplest flat direction VEV.  
{}From gauge invariance we find 
that the only mass terms appearing are for 
the hidden sector $\mbf 4$ and $\bar {\mbf 4}$ fields.\footnote{No 
$\mbf 4$ and $\bar {\mbf 4}$ 
fields of $SU(4)_c$ mass terms of this type appear either.}
For these fields all gauge invariant mass terms are also allowed by
picture--changed worldsheet charge invariance, being of worldsheet 
$[2_R, 2_R, 2_R]$ class.
These mass terms are all
sixth order and have coupling constants of equal magnitude. The latter
results from the $\IZ_2 \times \IZ_2$ worldsheet symmetry still present 
among these terms. The specific terms are:
\beqn
&&(\H{1}\bH{1'}+ \H{2}\bH{2'}) \vev{      \cL{L2}\cL{L3}      \cL{R2}\cL{R3}}
  \nolabel\\
&&(\H{1}\bH{3'}+ \H{2}\bH{4'}) \vev{      \cL{L2}\cL{L3}\cL{R1}      \cL{R3}},
  \nolabel\\
&&(\H{1}\bH{5'}+ \H{2}\bH{6'}) \vev{      \cL{L2}\cL{L3}\cL{R1}\cL{R2}      },
  \nolabel\\
&&(\H{3}\bH{1'}+ \H{4}\bH{2'}) \vev{\cL{L1}    \cL{L3}      \cL{R2}\cL{R3}}, 
  \nolabel\\
&&(\H{3}\bH{3'}+ \H{4}\bH{4'}) \vev{\cL{L1}    \cL{L3}\cL{R1}    \cL{R3}},
  \nolabel\\
&&(\H{3}\bH{5'}+ \H{4}\bH{6'}) \vev{\cL{L1}    \cL{L3}\cL{R1}\cL{R2}      },
  \nolabel\\
&&(\H{5}\bH{1'}+ \H{6}\bH{2'}) \vev{\cL{L1}\cL{L2}          \cL{R2}\cL{R3}},
  \nolabel\\
&&(\H{5}\bH{3'}+ \H{6}\bH{4'}) \vev{\cL{L1}\cL{L2}    \cL{R1}     \cL{R3}},
  \nolabel\\
&&(\H{5}\bH{5'}+ \H{6}\bH{6'}) \vev{\cL{L1}\cL{L2}    \cL{R1}\cL{R2}     }. 
\label{heq}
\eeqn

Assuming identical phase factors for each of these terms, 
the eigenstates and mass eigenvalues for the 
$SU(4)_{H_1}$ mass matrix are 
\beqn
&& \H{a}= \sixth (2\H{2}   - \H{4}   - \H{6}),\quad 
\bH{a}= \sixth (2\Hb{2'} - \Hb{4'} - \Hb{6'}); \quad
   M^2_a= 0 \label{hma}\\  
&&\H{b}= \half ( \H{4} - \H{6}),\quad \bH{b}= \half ( \Hb{4'} - \Hb{6'}); \quad
   M^2_b= 0 \label{hmb}\\   
&& \H{c}= \third (\H{2}   + \H{4}   + \H{6}),\quad 
\bH{c}= \third
 (\Hb{2'} + \Hb{4'} + \Hb{6'}); \quad
   M^2_c\approx \hund \MP^2
   \, . \label{hmc}  
\eeqn
Thus, for $SU(4)_{H_1}$ we have 
$N_c= 4$ and $N_f= 2$. This yields  
$\beta = -10$ and results in an $SU(4)_{H_1}$ 
condensation scale
\beqn
\Lambda_{H_1} = {\rm e}^{-15.8}\MP \sim 3\times 10^{11} \,\, {\rm GeV}. 
\label{consca2}
\eeqn

The $SU(4)_{H_2}$ eigenstates and eigenvalues may be converted from 
those of $SU(4)_{H_1}$ by exchanging field subscripts 
$(1\leftrightarrow 2)$, 
$(3\leftrightarrow 4)$, and 
$(5\leftrightarrow 6)$ and adding one massless ${\mbf 6}- {\mbf 6}$ 
vector pair.
The additional vector pair slightly lowers the $SU(4)_{H_2}$ condensation
scale to around 
\beqn
\Lambda_{H_2} = {\rm e}^{-17.5}\MP \sim 6\times 10^{10} \,\, {\rm GeV}. 
\label{consca3}
\eeqn
Hidden sector condensation scale of this order, together with the
hidden sector matter condensates and the superpotential terms eq.
(\ref{heq}), can indeed induce supersymmetry breaking at a
phenomenologically viable scale \cite{fh1,mshsm}. As has
been argued above, the LRS string model discussed here does not give
rise to a phenomenologically viable vacuum. The new interesting
feature of our LRS string model (\ref{model1},\ref{phasesmodel1})
is the fact that supersymmetry is hierarchically broken in the vacuum
because of the necessity to utilize non--Abelian VEVs. 

\section{Discussion}

The free fermionic heterotic string models are the only known 
string models that reproduced the Minimal Supersymmetric 
Standard Model in the effective low energy field theory. 
The important characteristics of such models is the 
generation of solely the MSSM spectrum in the effective low 
energy field theory, as well as the canonical normalization of the
weak--hypercharge, and the general GUT embedding of the Standard Model
spectrum, like the $SO(10)$ embedding. These characteristics
are well motivated by the structure of the Standard Model itself, 
as well as the MSSM gauge coupling unification prediction. 
This should be contrasted with the case of type I string models
in which one does not obtain the compelling GUT picture, but in which
the Standard Model gauge group arises from a product of $U(n)$ groups
\cite{ibaneztypeI}. The phenomenological success of the free fermionic 
models gives rise to the possibility that the true string vacuum 
lies in the vicinity of these models and justifies the continued
efforts to understand the general properties of these models. 

In this paper we examined the existence of a supersymmetric vacuum 
in the LRS free fermionic heterotic string models. This class
of models exhibits new features with respect to the anomalous 
$U(1)$ symmetry. In ref. \cite{lrsmodels}, in contrast to the
case of the FSU5, SLM and PS free fermionic string models, 
the existence of some LRS string models with vanishing $U(1)_A$
was demonstrated, whereas some LRS models did contain an anomalous
$U(1)$ symmetry. In this paper we observed the absence of 
singlet flat directions in the LRS string model which contained
an anomalous $U(1)$. This observation prompted the exciting
possibility that a supersymmetric vacuum does not exist in this
model, and that a non--vanishing Fayet--Iliopoulos $D$--term is
generated at one--loop in string perturbation theory, whereas 
the one--loop partition function still vanishes because
of the one--loop fermion--boson degeneracy. However, we demonstrated
that a $D$--flat vacuum, as well as $F$--flat to all finite orders, 
does exist if some non--Abelian fields in the massless string spectrum
obtain a non--vanishing VEV. This is the first instance in the
study of the realistic free fermionic models in which non--Abelian
VEVs are enforced by the requirement of a stable vacuum rather 
than by other phenomenological considerations. This situation 
still raises interesting prospects for the issue of supersymmetry
breaking for the following reasons. It has been shown that if
one utilizes solely singlet VEVs in the cancellation of the 
Fayet--Iliopoulos $D$--term then the vacuum can remain supersymmetric
to all orders. In contrast it has been argued in the past that
utilizing non--Abelian VEVs necessarily results in generation of
superpotential mass terms via hidden sector matter condensates that
results in hierarchical supersymmetry breaking \cite{fh1,mshsm}.
The study of these issues in closely related models is therefore
of further interest and will be reported in future publications.

\section{Acknowledgments}

We would like to thank Ignatios Antoniadis, Elias Kiritsis,
and Joel Walker for useful discussions. 
This work is supported in part by 
a PPARC advanced fellowship (AEF); and by DOE Grant
No.\  DE--FG--0395ER40917 (GBC).

\newpage


\vfill\eject

{{\oddsidemargin  10.5pt \evensidemargin  10.5pt
\textheight  612pt \textwidth  432pt
\headheight  12pt \headsep  20pt
\footheight  12pt \footskip  40pt

\bibliographystyle{unsrt}
}}


\textwidth=7.5in
\oddsidemargin=-18mm
\topmargin=-5mm
\renewcommand{\baselinestretch}{1.3}
\smallskip

\begin{table}
{\rm \large\bf Left--Right~Symmetric~Model~1~Fields}
\begin{eqnarray*}
\begin{tabular}{|c|c|c|rrrrrrrr|c|}
\hline

  $F$ & SEC & $(C;L;R)$ 
   & $Q_A$ & $Q_{1^\prime}$ & $Q_{2^\prime}$ & $Q_{3^\prime}$ & $Q_{4^\prime}$ &
             $Q_{5^\prime}$ & $Q_{7^{\prime\prime}}$ & $Q_{8^\prime}$ 
   & $SU(4)_{H_{1;2}}$ \\

\hline

  $Q_{L_1}$ &                          & $(4,2,1)$
     & -2  &  2  &  2  &   4  & -2  & -2 &  0  &  8  & $(1,1)$\\
  $Q_{R_1}$ & $\mb_1\oplus$ & $({\bar4},1,2)$
     & -2  & -2  & -2  &  -4  & -2  & -2 &  0  & -8 &  $(1,1)$\\
  $L_{L_1}$ &  $\mb_1+\mzeta+2\mgamma$ & $(1,2,1)$
     & -2  &  2  &  2  & -20  & -2  & -2 &  0  &  0 & $(1,1)$ \\
  $L_{R_1}$ & & $(1,1,2)$
     & -2  & -2  & -2  &  20  & -2  & -2 &  0  &  0 & $(1,1)$ \\
  $\cL{L_{1}}$  &         & $(1,2,1)$
     &  2 &   2  &  2  &   4  & -2  &  2 &  0  &-32 & $(1,1)$ \\
  $\cL{R_{1}}$  &         & $(1,1,2)$
     &  2 &  -2  & -2  &  -4  &  2  &  2 &  0  & 32 & $(1,1)$ \\

\hline

  $Q_{L_2}$ &         			& $(4,2,1)$
     & -2 & -2 &  2 &   4 &  2 & -2 &  0 &  8 & $(1,1)$   \\
  $Q_{R_2}$ & $\mb_2\oplus$ & $({\bar4},1,2)$ 
     & -2 &  2 & -2 &  -4 &  2 & -2 &  0 & -8 & $(1,1)$   \\
  $L_{L_2}$ & $\mb_2+\mzeta +2\mgamma$ & $(1,2,1)$
     & -2 & -2 &  2 & -20 &  2 & -2 &  0 &  0 & $(1,1)$   \\
  $L_{R_2}$ &         & $(1,1,2)$
     & -2 &  2 & -2 &  20 &  2 & -2 &  0 &  0 & $(1,1)$   \\
  $\cL{L_{2}}$  &        & $(1,2,1)$
     &  2 & -2 &  2 &   4 & -2 &  2 &  0 &-32 & $(1,1)$   \\
  $\cL{R_{2}}$  &        & $(1,1,2)$
     &  2 &  2 & -2 &  -4 &  2 &  2 &  0 & 32 & $(1,1)$   \\

\hline

  $Q_{L_3}$ &  & $(4,2,1)$
     &  -2 &  0 &  -4 &  4 &  0 &  4 &  0 &  8 & $(1,1)$     \\
  $Q_{R_3}$ & $\mb_3\oplus$ & $({\bar4},1,2)$
     &  -2 &  0 &   4 & -4 &  0 &  4 &  0 & -8 & $(1,1)$     \\
  $L_{L_3}$ & $\mb_3+\mzeta +2\mgamma$ & $(1,2,1)$
     &  -2 &  0 &  -4 &-20 &  0 &  4 &  0 &  0 & $(1,1)$     \\
  $L_{R_3}$ & & $(1,1,2)$
     &  -2 &  0 &   4 & 20 &  0 &  4 &  0 &  0 & $(1,1)$     \\
  $\cL{L_{3}}$  &         & $(1,2,1)$
     &   2 &  0 &  -4 &  4 &  0 & -4 &  0 & -32 & $(1,1)$    \\
  $\cL{R_{3}}$  &         & $(1,1,2)$
     &   2 &  0 &   4 & -4 &  0 & -4 &  0 &  32 & $(1,1)$    \\

\hline

  $\Phi_1$          &         & $(1,1,1)$
     &  0 &  0 &  0 &  0 &  0 &  0 &  0 &  0 & $(1,1)$        \\
  $\Phi_2$          &         & $(1,1,1)$
     &  0 &  0 &  0 &  0 &  0 &  0 &  0 &  0 & $(1,1)$        \\
  $\Phi_3$          &         & $(1,1,1)$
     &  0 &  0 &  0 &  0 &  0 &  0 &  0 &  0 & $(1,1)$        \\
  $\Phi_{12}$       & Neveu-  & $(1,1,1)$
     &  0 & -8 &  0 &  0 &  0 &  0 &  0 &  0 & $(1,1)$        \\
  ${\bar\Phi}_{12}$ & Schwarz & $(1,1,1)$
     &  0 &  8 &  0 &  0 &  0 &  0 &  0 &  0 & $(1,1)$        \\
  $\Phi_{23}$       &         & $(1,1,1)$
     &  0 &  4 &-12 &  0 &  0 &  0 &  0 &  0 & $(1,1)$        \\
  ${\bar\Phi}_{23}$ &         & $(1,1,1)$
     &  0 & -4 & 12 &  0 &  0 &  0 &  0 &  0 & $(1,1)$        \\
  $\Phi_{31}$       &         & $(1,1,1)$
     &  0 & -4 & -12&  0 &  0 &  0 &  0 &  0 & $(1,1)$        \\
  ${\bar\Phi}_{31}$ &         & $(1,1,1)$
     &  0 &  4 &  12&  0 &  0 &  0 &  0 &  0 & $(1,1)$        \\

\hline

$\D{3}$& & $(4,1,1)$
     & 0 &  0 &  4 & 8 &  0 &  0 &  0 &  -24 & $(1,1)$        \\

$\Db{3}$& & $({\bar 4},1,1)$
     & 0 &  0 &  -4 & -8 &  0 &  0 &  0 &  24 & $(1,1)$   \\

${\phi}_{\alpha\beta}$& & $(1,1,1)$
     & 0 &  0 & -12 & 0  & 0 &  0 &  0 &  0 & $(1,1)$        \\

${\bar \phi}_{\alpha\beta}$& $\xi\equiv\mS+\mb_1+\mb_2+$  & $(1,1,1)$
     & 0 &  0 &  12 & 0  & 0 &  0 &  0 &  0 & $(1,1)$        \\

${\phi}_{1}$& $\malpha+\mbeta$ & $(1,1,1)$
     &  0 & 4 &  0 & 0 & 0 &  0 &  0 &  0 & $(1,1)$        \\

${\bar \phi}_{1}$& $\oplus$ & $(1,1,1)$
     &  0 & -4 &  0 & 0 & 0 &  0 &  0 &  0 & $(1,1)$        \\

${\phi}_{2}$& $\xi+\zeta$ & $(1,1,1)$
     &  0 & 4 &  0 & 0 &  0 &  0 &  0 &  0 & $(1,1)$        \\

${\bar \phi}_{2}$ &  & $(1,1,1)$
     &  0 &  -4 &  0 & 0 &  0 &  0 &  0 &  0 & $(1,1)$        \\

$\S{8}$ &  & $(1,1,1)$
     & 0 &  0 & 4 &  -16 &  0 &  0 &  0 &  -32 & $(1,1)$        \\

$\bS{8}$ & & $(1,1,1)$
     &  0 & 0 &  -4 & 16 &  0 &  0 & 0 &  32 & $(1,1)$        \\

\hline
\end{tabular}
\nolabel
\end{eqnarray*}
Table 1: {\it Model 1 fields}.
\end{table}


\begin{flushleft}
\begin{table}
{\rm \large\bf Left--Right~Symmetric~Model~1~Fields~Continued}
\begin{eqnarray*}
\begin{tabular}{|c|c|c|rrrrrrrr|c|}
\hline

  $F$ & SEC & $(C;L;R)$ 
   & $Q_A$ & $Q_{1^\prime}$ & $Q_{2^\prime}$ & $Q_{3^\prime}$ & $Q_{4^\prime}$ &
             $Q_{5^\prime}$ & $Q_{7^{\prime\prime}}$ & $Q_{8^\prime}$ 
   & $SU(4)_{H_{1;2}}$ \\

\hline

  $\D{1}$ & & $ (4,1,1)$
     &  0 &   2  &  2 &   4 &  0 &  0 &  -8 &  8 & $(1,1)$     \\

  $\Db{1}$        &                      &   $ ({\bar 4},1,1)$
     &  0 &  -2 &  -2 &  -4 &  0 &  0 &   8 & -8 & $(1,1)$    \\

  $\S{1}$        & $\xi\equiv\mS+ \mb_2+\mb_3+$        &   $(1,1,1)$
     &  0 &  -2 &  6 &  -12 &  0 &  0 &  8 & 16 & $(1,1)$        \\

  $\bS{1}$ &  $~\mbeta+\mgamma~$ &   $(1,1,1)$
     &  0 & 2 &  -6 &  12 &  0 &  0 &  -8 & -16 & $(1,1)$         \\

  $\S{2}$ &       $\oplus$          &   $(1,1,1)$
     &  0 &  2 &  -6 &  -12 &  0 &  0 &  8 & 16 & $(1,1)$        \\

  $\bS{2}$ & $\xi+\zeta+2\gamma$ &  $(1,1,1)$
     &  0 &  -2 & 6 & 12 &  0 &  0 &  -8 & -16 & $(1,1)$        \\

  ${\cH{1}}$ &                             &   $(1,1,1)$
     &  0 & 2 &  2 &  -8 &  0 &  0 &  0 & -16 & $(1,6)$         \\

 ${\cHb{1}}$ &                      &   $(1,1,1)$
     &  0 &  -2 & -2 & 8 &  0 &  0 &  0 & 16 & $(1,6)$ \\

\hline

$\D{2}$ & & $(4,1,1)$
     &  0 &  -2 & 2 & 4 & 0 &  0 &  -8 & 8 & $(1,1)$        \\

$\Db{2}$ &  & $({\bar 4},1,1)$
     &  0 & 2 &  -2 & -4 &  0 & 0 & 8  &  -8 & $(1,1)$        \\

$\S{3}$ &  $\xi\equiv\mS+\mb_1+\mb_3+$ & $(1,1,1)$
     &  0 &   2 &   6 &  -12 &  0 &  0 &   8 &  16 & $(1,1)$        \\

$\bS{3}$ & $\malpha+\mgamma$     & $(1,1,1)$
     &  0 &  -2 &  -6 &   12 &  0 &  0 &  -8 &  -16 & $(1,1)$        \\

$\S{4}$ &  $\oplus$  & $(1,1,1)$
     &  0 &  -2 &  -6 &  -12 &  0 &  0 &   8 & 16 & $(1,1)$        \\

${\bar S}_{4}$ & $\xi+\zeta+2\mgamma$ & $(1,1,1)$
     &  0 &  2 &  6 &  12 &  0 &  0 &  -8 &  -16 & $(1,1)$         \\

${\cH{2}}$ & & $(1,1,1)$
     &  0 &  -2 &  2 &  -8 &  0 & 0 &  0 &  -16 & $(1,6)$        \\

${\cHb{2}}$ & & $(1,1,1)$
     &  0 &  2 &  -2 &  8 & 0 &  0 &  0 &  16 & $(1,6)$         \\

\hline

$H_{1}$ & $\xi\equiv\mS+\mb_2+\mb_3+$ & $(1,1,1)$
     &  -4 & -2 & -2 &   2 &  2 &   2 & -4 & -16 & $(1,4)$        \\

${\bar H}_{1'}$ & $\malpha+2\mgamma$ & $(1,1,1)$
     &  -4 &  2 &  2 &  -2 &  2 &   2 &  4 &  16 & $(1,{\bar4})$        \\

$ H_{2}$ & $\oplus$ & $(1,1,1)$
     &  -4 & -2 & -2 &   2 &  2 &   2 &  4 & -16 & $(4,1)$        \\

${\bar H}_{2'}$ & $\xi+\zeta$ & $(1,1,1)$
     &  -4 &  2 &  2 &  -2 &  2 &   2 & -4 &  16 & $({\bar 4},1)$        \\

\hline

$H_{3}$& $\xi\equiv\mS+\mb_1+\mb_3+$ & $(1,1,1)$
     & -4 &  2 & -2 &  2 &  -2 &  2 &  -4 &  -16 & $(1,4)$        \\

${\bar H}_{3'}$ &  $\malpha+2\mgamma$ & $(1,1,1)$
     & -4 &  -2 &  2 &  -2 &  -2 &  2 &  4 &  16 & $(1,{\bar4})$        \\

$H_{4}$ & $\oplus$& $(1,1,1)$
     &  -4 &  2 &  -2 & 2 &  -2 &  2 &  4 &  -16 & $(4,1)$        \\

${\bar H}_{4'}$ &$\xi+\zeta$ & $(1,1,1)$
     &  -4 & -2 &  2 & -2 &  -2 &  2 &  -4 &  16 & $({\bar 4},1)$        \\

\hline

$H_{5}$& $\xi\equiv\mS+\mb_1+\mb_2+$ & $(1,1,1)$
     & -4 &  0 &   4 &  2 &  0 & -4 & -4 & -16 & $(1,4)$        \\

${\bar H}_{5'}$ &  $\malpha+2\mgamma$ & $(1,1,1)$
     & -4 &  0 &  -4 & -2 &  0 & -4 &  4 &  16 & $(1,{\bar4})$        \\

$H_{6}$ & $\oplus$ & $(1,1,1)$
     &  -4 & 0 &   4 &  2 &  0 & -4 &  4 & -16 & $(4,1)$        \\

${\bar H}_{6'}$ &  $\xi+\zeta+2\mgamma$ & $(1,1,1)$
     &  -4 &  0 & -4 & -2 &  0 & -4 & -4 &  16 & $({\bar4},1)$        \\

\hline

${S}_{5}$ &  & $(1,1,1)$
     & 0 &  0 & -8 & -16  & 0 &  0 &  0 &  -32 & $(1,1)$        \\

${\bar S}_{5}$ & & $(1,1,1)$
     &  0 & 0 & 8 & 16 & 0 &  0 &  0 &  32 & $(1,1)$        \\

${S}_{6}$ & $\mS+\zeta+2\mgamma$ & $(1,1,1)$
     &  0 & -4 &  4 & -16 & 0 &  0 &  0 &  -32 & $(1,1)$        \\

${\bar S}_{6}$ & & $(1,1,1)$
     &  0 &  4 &  -4 & 16 & 0 &  0 &  0 & 32 & $(1,1)$        \\

${S}_{7}$ & & $(1,1,1)$
     &  0 &  4 &   4 & -16 & 0 &  0 &  0 &  -32 & $(1,1)$        \\

${\bar S}_{7}$ & & $(1,1,1)$
     &  0 & -4 &  -4 &  16 & 0 &  0 &  0 &   32 & $(1,1)$        \\

\hline

\end{tabular}
\nolabel
\end{eqnarray*}
{Table 1: \it Model 1 fields continued}.
\end{table}
\end{flushleft}

\begin{table}[!ht]
\begin{flushleft}
\begin{tabular}{|l|r|rrrrrrrr|r|}
\hline
Basis  &$\QA$& $\S{5}/\bS{5}$  
                   & $\Db{3}/\D{3}$    
                         & $\Db{2}/\D{2}$
                               & $\cH{2}/\cHb{2}$   
                                     & $\cH{1}/\cHb{1}$   
                                           & $\bH{4'}$ 
                                                & $\bH{2'}$
                                                     &    &     \\
\hline
$\Dm{ 1}$& 12&   0 & -1  &  1  & -2  &  3  &  0 & -2 &  2 & $\cL{R1}$ \\
$\Dm{ 2}$& 12&  -1 &  3  & -1  &  0  &  1  &  0 & -2 &  2 & $\cL{L2}$ \\
$\Dm{ 3}$& 12&  -1 &  3  & -1  &  0  &  1  & -2 &  0 &  2 & $\cL{L1}$ \\
$\Dm{ 4}$& 12&   0 & -1  & -1  &  2  & -1  & -2 &  0 &  2 & $\cL{R2}$ \\
$\Dm{ 7}$& 24&  -1 &  4  & -4  & -2  &  2  & -4 & -4 &  4 & $\Q{R3}$ \\
$\Dm{ 5}$& 24&  -3 &  4  & -4  &  2  &  6  & -4 & -4 &  4 & $\Q{L3}$ \\
$\Dm{ 6}$& 24&  -1 &  8  & -4  &  2  &  6  & -4 & -4 &  4 & $\L{R3}$ \\
$\Dm{ 8}$& 24&  -3 &  0  & -4  & -2  &  2  & -4 & -4 &  2 & $\L{L3}$ \\ 
\hline
$\Dm{ 9}$&-12&   0 & -1  &  1  &  0  & -1  &  2 &  0 &  2 & $\Q{R2}$ \\
$\Dm{10}$&-12&   1 & -3  &  1  & -4  & -1  &  2 &  0 &  2 & $\L{L2}$ \\
$\Dm{11}$&-12&   1 & -1  &  1  & -2  &  1  &  2 &  0 &  2 & $\Q{L2}$ \\
$\Dm{13}$&-12&   0 &  1  &  1  &  2  &  1  &  2 &  0 &  2 & $\L{R2}$ \\
$\Dm{12}$&-12&   0 &  1  &  1  &  2  &  1  &  0 &  2 &  2 & $\L{R1}$ \\
$\Dm{14}$&-12&   1 & -3  &  1  &  0  & -5  &  0 &  2 &  2 & $\L{L1}$ \\
$\Dm{15}$&-12&   0 & -1  &  1  &  0  & -1  &  0 &  2 &  2 & $\Q{R1}$ \\
$\Dm{16}$&-12&   1 & -1  &  1  &  2  & -3  &  0 &  2 &  2 & $\Q{L1}$ \\
$\Dm{17}$&-24&   1 & -3  &  1  &  0  & -1  &  2 &  2 &  2 & $\bH{5'}$ \\
$\Dm{18}$&-24&   1 & -1  &  3  &  0  & -3  &  2 &  2 &  2 & $\H{5}$ \\
$\Dm{19}$&-24&  -1 &  0  &  4  &  2  & -2  &  4 &  4 &  4 & $\L{L3}$ \\
$\Dm{20}$&-24&   5 & -8  &  4  & -2  & -6  &  4 &  4 &  4 & $\L{R3}$ \\
$\Dm{21}$&-48&   1 & -6  &  6  &  2  & -4  &  4 &  4 &  4 & $\bH{6'}$ \\
$\Dm{22}$&-48&   3 & -2  &  2  & -2  & -4  &  4 &  4 &  4 & $\H{6}$ \\
\hline
$\Dm{40}$&  0&  -3 &  4  &  0  & -2  &  6  &  0 & -4 &  4 & $\H{1}$ \\
$\Dm{37}$&  0&   1 &  0  & -4  & -2  &  2  &  0 & -4 &  4 & $\bH{1'}$ \\
$\Dm{41}$&  0&  -3 &  4  &  0  &  6  & -2  & -4 &  0 &  4 & $\H{3}$ \\
$\Dm{29}$&  0&   1 &  0  & -4  & -2  &  2  & -4 &  0 &  4 & $\bH{3'}$ \\
$\Dm{34}$&  0&  -1 &  2  & -2  & -2  &  4  &  0 & -2 &  2 & $\H{2}$ \\
$\Dm{39}$&  0&  -1 &  2  & -2  &  2  &  0  &  0 & -2 &  2 & $\H{4}$ \\
\hline
$\Dmv{32}$& 0&   0 &  0  &  0  & -2  &  2  &  0 &  0 &  1 & $\P{12}$  \\ 
$\Dmv{35}$& 0&  -1 &  0  &  0  &  2  &  0  &  0 &  0 &  1 & $\P{23}$  \\
$\Dmv{28}$& 0&  -1 &  0  &  0  &  0  &  2  &  0 &  0 &  1 & $\P{31}$  \\ 
$\Dmv{31}$& 0&  -1 &  0  &  0  &  1  &  1  &  0 &  0 &  1 & $\S{8}$ \\ 
$\Dmv{25}$& 0&   0 &  0  &  0  & -1  & -1  &  0 &  0 &  1 &$\p{\alpha\beta}$\\ 
$\Dmv{33}$& 0&   0 &  0  &  0  & -1  &  1  &  0 &  0 &  1 & $\pb{1,2}$\\ 
$\Dmv{30}$& 0&   0 & -1  & -1  & -1  &  1  &  0 &  0 &  1 & $\S{4}$ \\ 
$\Dmv{27}$& 0&   0 & -1  & -1  &  0  &  0  &  0 &  0 &  1 & $\S{2}$ \\
$\Dmv{36}$& 0&   1 & -1  & -1  & -1  & -1  &  0 &  0 &  1 & $\S{3}$\\ 
$\Dmv{24}$& 0&   1 & -1  & -1  & -2  &  0  &  0 &  0 &  1 & $\S{1}$ \\
$\Dmv{23}$& 0&   0 &  0  &  0  &  0  & -2  &  0 &  0 &  1 & $\S{7}$ \\ 
$\Dmv{38}$& 0&   0 &  0  & -2  &  0  &  0  &  0 &  0 &  1 & $\S{6}$ \\
$\Dmv{26}$& 0&   0 &  0  & -1  & -1  &  1  &  0 &  0 &  1 & $\Db{1}$   \\ 
\hline \hline
\end{tabular}
\end{flushleft}
{Table 2: \it $D$--Flat Direction Basis Set for Model 3.}
\end{table}

\end{document}